\documentclass[ amssymb,amsmath,notitlepage]{revtex4-1}

\usepackage{dcolumn}
\usepackage{bm}
\usepackage[T1]{fontenc}
\usepackage{lmodern}

\usepackage{color}
\usepackage{graphicx}
\usepackage{epstopdf}
\usepackage{hyperref}

\begin{document}

\title{On the pressure exerted by a bundle 
of independent living filaments}

\author{Jean-Paul Ryckaert}
\email{jryckaer@ulb.ac.be}
\author{Sanoop Ramachandran}
\affiliation{ 
 Physique des Polym\`eres, Universit\'e Libre de Bruxelles,\\
Campus Plaine, CP 223, B-1050 Brussels, Belgium
}%

\date{\today}

\begin{abstract}
The properties of a bundle of grafted semi-flexible 
living filaments in ideal solution facing an obstacle wall, under supercritical 
conditions, are explored.
For this purpose, we make use of the discrete wormlike chain model
characterized by the monomer size $d$, a size dependent contour length 
$L_{\rm c}$ and a persistence length $l_{\rm p}$.
The calculation of the equilibrium filament size distribution and the average equilibrium force require the knowledge of the wall effect on the 
single filament partition function of any size, which can be computed by Metropolis Monte-Carlo methods.
The force exerted by a living filament on 
a fixed wall turns out to be the weighted average of the dead grafted 
filament forces computed for sizes hitting the wall, multiplied
by the probability of occurrence of the corresponding filament size. 
As the distance to the wall is varied, the resultant 
force shows large variations whose amplitude decrease with increasing 
gap sizes and/or with decreasing persistence length.
Also, its average over a gap interval of precise size $d$ gives an 
average force close to what is expected by the ratchet model 
for actin growth against a wall. 
The osmotic pressure exerted by $N_f$ filaments is the average equilibrium force per filament
times the grafting surface density.
\end{abstract}

\maketitle

\section{Introduction. \label{sec:intro}}

In eukaryotic cells, the various (de)polymerization reactions 
of cytoskeletal filaments such as actin and microtubules are coupled to various physical processes associated with cell motility.  
Some examples are the filopodia or lamellopodes whose underlying
actin network structures grow against the cell membrane and hence develop 
protrusive forces and associated work. 
Similarly, the mobility of the pathogenic bacterium, lysteria, inside the 
infected cell results from a force originating from the build up
of an actin network within a host cell at the rear of the bacterium membrane, which pushes forward the lysteria~\cite{b.Howard}. 
Focusing on the actin case in supercritical conditions, where the polymerization 
steps dominate, the direct or indirect measurement of these polymerizing 
forces for a controlled number of filaments have shown that they are of the 
order of a few pN per 
filament~\cite{KP.04, Berrot.07, Dogterom.07, Brangbour.11}.

Statistical mechanics investigations on these polymerization forces 
require an explicit consideration of chemical reactions. 

In his seminal papers~\cite{Hill.81,Hill.82}, T. L. Hill investigated the 
influence of tension or compression, within a straight polymerizing filament 
tethered at both ends, on the insertion probability of a monomer 
within the filament. 
If $\rho_{1c}$ represents the monomer solution density in chemical equilibrium 
with monomers in the tension-free filament (zero internal force), he 
established that for an arbitrary monomer density $\rho_1$, the equilibrium 
will be attained when the filament internal force $F$ (taken here as 
positive when compressive) reaches a value given by
\begin{equation}
F=\frac{k_{\rm B}T}{d}  \ln{\left(\frac{\rho_1}{\rho_{1c}}\right)},
\label{eq:force}
\end{equation}
where $d$ is the filament incremental contour length due to the insertion 
of a monomer in the structure. As an example, $d$ is taken to be 2.7 nm for actin which is the 
half of the globular-actin (G-actin) diameter since filamentous-actin 
(F-actin) is a double stranded filament. 
For filaments with at least one free end, the equilibrium polymer 
model~\cite{b.Howard} establishes a critical concentration $\rho_{1c}$ for 
which polymerization and depolymerization reactions proceed at equal rates 
at the actin filament end. 
In presence of a compressive force $F$, the mechanical energy of the ($n$+1)-mer 
exceeds the one of the $n$-mer by $Fd$. 
Hence the equilibrium constant is displaced towards depolymerization. 
Equation (\ref{eq:force}) is recovered via another interpretation of the 
critical density $\rho_{1c}$ when $F$ is the compressive force exerted at the tip of the filament which is needed to get 
chemical equilibrium at density $\rho_{1}$. 
Such a situation is relevant when a filament in supercritical conditions,
 grafted to an actin network at one side, is compressed by a membrane wall that 
its free ends encounters as it elongates. 
To interpret how actin monomers could intercalate in between the tip of the 
pushing filament and the wall (e.g. the bacterium membrane), Peskin et 
al. \cite{Pes.93} suggested a Brownian ratchet model whereby the rigid filaments 
can grow against a wall subjected to an opposite load force by filling the 
sufficiently large gaps produced by the wall thermal fluctuations. 
This mechanism could not be validated as such experimentally, suggesting a 
close but alternative model, the elastic Brownian ratchet model whereby bending 
undulations of the filaments create the needed gap \cite{Mol.96}. 
It must be stressed that it is generally believed that ultimate understanding 
for the force production in biological situations also requires a macroscopic 
viscoelastic treatment of the actin gel from which some individual filaments 
originate to transmit forces~\cite{Joanny.09}.

A relatively simple actin network is found at the tip of filopodial structures. 
A bundle of grafted filaments emerges from a network of cross linked parallel 
filaments and hits a membrane. 
Such a network has been the object of theoretical 
studies~\cite{Joanny.11,Krawczyk.11} assuming a dynamical stochastic model for 
a set of parallel rigid living filaments pressing on a mobile rigid piston, 
always located at the level the longest filament of the bundle. 
The stalling force for a set of $N$ non-interacting filaments is found to be 
$N$ times the force per filament given by Eq. (\ref{eq:force}). 
In presence of lateral attractive interactions between filaments, the 
equilibrium force exerted by the bundle on the wall is enhanced by an 
additive contribution N $\epsilon_l$, where $\epsilon_l$ is the attraction 
energy gain per unit length~\cite{Krawczyk.11}. 
The stalling force needed to stop further polymerization of a network of 
parallel filaments in supercritical conditions was probed experimentally 
in vitro in \cite{Dogterom.07} for a bundle of $N_f \approx 10$ filaments, but 
the measured force was found to be considerably lower than $N_f$ 
times the single filament force Eq.(\ref{eq:force}), an experimental result 
whose interpretation is still under debate~\cite{tsekouras.11,mogilner.09}.

Seeking for a more microscopic description of a wall blocking a set of 
growing filaments, we have recently started a series of mesoscopic simulations 
of a bundle of grafted living biofilaments with persistence length $l_{\rm p}$, 
in chemical equilibrium with explicit free monomers~\cite{RR.13}, based on 
the living filament model proposed earlier~\cite{Caby.12} in which intermolecular forces are
explicitly taken into account. 
Each filament is modelled as a discrete chain of freely rotating stiff bonds, 
with a bending penalty term to adjust the persistence length. 
Its infinitely stiff bond limit corresponds to a discrete wormlike chain 
model characterized by three length scales, the contour length $L_c$, the persistence length $l_p$
and the step length $d$ corresponding to the contour length change per incorporated or leaving single 
monomer. 
Articulation points of the chain coinciding with Lennard-Jones centres of 
forces describe intermolecular interactions. 
In the simulated slit pore, the filaments anchored to one planar wall are 
growing in supercritical conditions and hit the opposing parallel wall located 
at a distance $L$ from the first wall. 
An equilibrium situation is established when filament-wall forces become able 
to stop further growth of these filaments (in the average). 
These simulations are based on a formal Hamiltonian describing a set of $N_1$ 
free monomers and a set of $N_f$ filaments of varying size in contact with 
two parallel walls. 
While following the system composed of free monomers, grafted filaments and 
mesoscopic heat bath solvent particles by Molecular Dynamics, 
explicit Monte-Carlo moves do model the chemical (de)polymerizing steps in a 
way satisfying micro-reversibility~\cite{Caby.12}. 
In this way, our simulations sample a non-ideal reactive canonical ensemble 
for the grafted and confined bundle system. 
We find that the distribution of filaments lengths 
is exponential for sizes for 
which no direct interaction with the obstacle wall is possible. 
The wall presence induces the chemical species, corresponding to filaments 
with contour length longer than the gap size $L$, to bend and thus to store 
some effective compressional energy which more or less strongly, decreases their
probability of occurrence at equilibrium. 
Hence, the size distribution decreases to zero over a few steps provided 
the supercriticality (chemical affinity) is appropriately limited to avoid 
lateral escape of very long filaments.

To interpret correctly the non-ideal effects related to
monomer-monomer inter-filament interactions in the simulations, in 
particular the observed decrease of the pressure exerted by the 
filaments on the opposite wall~\cite{RR.13}, it is important to have a detailed understanding 
of the already complex ideal solution case where the 
filaments are independent. 
Such an ideal bundle system, in chemical equilibrium with free monomers, 
presents a rich set of properties which could not be developed in 
reference~\cite{RR.13} and which is therefore the main theme covered by the present article. 
Two essential ingredients appear in the theoretical treatment. 
First, the free monomer reduced density $\hat{\rho}_1=K_0 \rho_1$, where $K_0$ is the 
(de)polymerizing equilibrium constant of a free-end in the ideal reference 
state imposes the degree of supercriticality 
(or possibly the degree of subcriticality) which is reflected by the 
characteristic length scale $1/\ln{\hat{\rho}_1}$ associated to the 
exponential distribution of the short filament lengths. 
The tail of the distribution corresponding to long filament sizes requires 
wall factors $\alpha_i(L;l_{\rm p},d)$. These are related to single filament partition functions 
of a grafted filament of size $i$ with fixed initial orientation, respectively 
in presence or in absence of the obstacle wall at a distance $L$ from the seed grafting wall.

The paper is organized as follows. 
Section \ref{sec:therm} provides the general thermodynamic description of the 
bundle of semi-flexible living filaments grafted at one wall, subject to 
single monomer (de)polymerization steps with free monomers of the bath, 
constrained within a slit pore with gap size $L\ll l_{\rm p}$. 
The specific statistical mechanics model of the filaments and the constraints 
or interactions imposed by the walls are formulated, allowing the
derivation of the partition functions of the filaments of arbitrary size. 
And explicit account of the chemical reactions leads to the formal 
reactive canonical ensemble which provides the fundamental link with 
the Helmholtz free energy. 
Within the ideal bundle approximation, the distribution of filament sizes 
and the osmotic pressure or the osmotic force per filament are then derived 
from the explicit reactive ensemble partition function. 
The section ends with considerations on the limits of our theoretical 
approach, namely the existence of a maximum supercriticality 
$\hat{\rho}_{1b}$ not to overcome to avoid any significant possibility for filaments to 
bend strongly over the gap distance $L$ and start polymerizing parallel to 
the obstacle wall.

In Section \ref{sec:Monte-Carlo}, we transform the formal expressions for the wall factor
into equivalent averages which turn out to be more appropriate for a 
single-filament/wall Monte-Carlo calculations, and we give details on their 
calculations for illustrative filament models, filament sizes and gap sizes. 
Most of the illustrations on the calculations of wall factors, filament size 
distributions at various reduced reduced free monomer densities and the 
corresponding equilibrium force for the living filaments, are concerned with 
the discrete wormlike chain model hitting a hard wall.
We also provide an example with the finite stiffness version of our filament 
model hitting a specific continuous repulsive wall, using the model treated 
by our non-ideal simulations~\cite{RR.13}.

Section \ref{sec:concl} summarizes the outcomes of our study and concludes on 
some implications of our results.

\section{The confined bundle system \label{sec:therm}}

\subsection{
Statistical Mechanics description of the bundle of grafted living filaments}

We are interested in a closed reacting ideal mixture at temperature $T$, 
having a fixed total number $N_t$ of monomers which can either be free 
monomers (as a G-actin-ATP complex) or integrated within self-assembled 
filaments of variable length (as an actin-ATP complex in F-actin). 
This system is enclosed in a slit pore of volume $V=AL$ with parallel walls 
of transverse area $A$ and gap width $L$ in which a fixed number $N_f$ of 
independent filaments with variable contour length $L_{\rm c}$ and fixed persistence 
length $l_{\rm p} \gg L$ are grafted normal to one of the walls. 
The individual filaments of this bundle are continuously growing or 
shrinking through single monomers reversible (de)polymerization reactions 
at their free ends, consuming/rejecting free monomers in/from the bath so 
that, at each reacting event, the filament contour length jumps by a step 
length $d$ representing the effective monomer size. 
This contour length $L_{\rm c}=(i-1)d$ is thus directly linked to the instantaneous 
total number of monomers $i$ in the particular filament. 
In supercritical conditions where polymerization dominates, these 
semi-flexible filaments will grow and hit the obstacle formed by the 
opposite wall of the pore as soon as $L_{\rm c}>L$. 
The global polymerization will be halted given the limited flexibility of 
the filaments and an equilibrium situation, characterized by an Helmholtz 
free energy $F(T,A,L,N_t,N_f)$, will be established for conditions we 
now establish.

The chemical reaction will be denoted as
\begin{equation}
A_{i-1} + A_1 \rightleftharpoons A_{i}      \ \ \ (3<i \leq z^{*})
\label{eq:reaction}
\end{equation}
where $A_i$ and $A_1$ represent respectively an anchored filament of size $i$ 
and a free monomer. 
This series of reactions is considered as limited to a size window going from 
a minimum filament size of three (to be considered as an effective permanent 
seed of the filament) up to a maximum size of 
\begin{equation}
z^{*}=\frac{\pi L}{2 d}.
\label{eq:z*}
\end{equation}
This upper limit corresponds to the size of a filament adopting a planar 
conformation of homogeneous curvature $1/L$ which, while satisfying end filament 
anchoring constraints and confinement, reorients its free end by 
$90$ degrees and becomes able to further grow parallel to the wall surface 
without any further bending energy penalty.
In the present work, we want to establish the compressive force originating 
from a bundle of relatively stiff filaments. 
We will restrict ourselves to thermodynamic conditions corresponding to a 
regime for which filaments of sizes approaching $z^{*}$ are practically 
unpopulated because of their too high bending energy with respect to thermal 
fluctuations. 
These conditions are made explicit in the following paragraphs.

Starting with the general differential form of the free energy for a 
non-reacting mixture of $N_1$ free monomers, $N_3$ anchored filaments of 
three monomers, $N_4$ anchored filaments of four monomers, $\dots$, one 
would have 
\begin{align}
dF= -SdT-p_N A dL&-p_T L dA+\mu_1 dN_1+\sum_{i=3}^{z^{*}} \mu_{i}dN_i,
\label{eq:dF1}
\end{align}
where the second and third reversible works terms imply the 
normal pressure ($p_N$) and tangential pressure ($p_T$) respectively. 
The other terms involve the chemical potentials $\mu_1,\mu_3,\dots$ of the 
different species. 
If we now relax the constraint of non-reactive mixture and impose the 
chemical equilibrium for all reactions in Eq.~(\ref{eq:reaction}), we have
\begin{equation}
\mu_{i-1} + \mu_1 = \mu_{i} \ \ \ (3 <i \leq z^{*}).
\label{eq:mu}
\end{equation}
At the same time, we impose that the relevant composition variables are 
$N_t$ and $N_f$, related to species composition variables by 
\begin{align}
N_t&= N_1+3 N_3+ 4 N_4+ 5 N_5+\ldots+ z^{*}N_{z^{*}},
\label{eq:constr1}\\
N_f&= N_3+ N_4+ N_5 + \ldots +N_{z^{*}}.
\label{eq:constr2}
\end{align}
Taking into account Eqs.~(\ref{eq:mu}),~(\ref{eq:constr1}) 
and~(\ref{eq:constr2}), Eq.~(\ref{eq:dF1}) becomes
\begin{align}
dF= -SdT-&p_N A dL-p_T L dA+\mu_1 dN_t+(\mu_3-3 \mu_1)dN_f,
\label{eq:dF2}
\end{align}
implying that the normal pressure $p_N$ and the free monomer chemical 
potential $\mu_1$ are given by 
\begin{align}
p_N&= -\frac{1}{A} \left(\frac{\partial F}{\partial L}\right)_{N_t,N_f,A,T},
\label{eq:pN}\\
\mu_1&= \left(\frac{\partial F}{\partial N_t}\right)_{N_f,A,L,T}.
\label{eq:mu1}
\end{align}

The link with statistical physics is given by $\beta F=- \ln{Q^{\rm RC}}$, 
where the reactive canonical ensemble expression 
$Q^{\rm RC}(N_t,N_f,L,A,T)$ relative to the anchored bundle system simplifies 
in the ideal solution approximation to \cite{b.Hill}
\begin{align}
Q^{\rm RC}=\sum_{N_1,N_3,.} 
\frac{q_1^{N_1}}{N_1!} q_3^{N_3}... 
q_z^{N_z}...q_{z+k^{*}}^{N_{z+k^{*}}} 
\frac{N_f!}{N_3! ..N_z!...N_{z+k^{*}}!},
\label{eq:rci}
\end{align}
where the sum runs over all distinct possible arrangements of $N_t$ monomers 
into various numbers of $N_1$ free monomers, $N_3$ filaments of length $i=3$, 
$N_4$ filaments of length $i=4$, $\ldots$ and $N_{z+k^{*}}$ filaments of the 
largest size allowed $z^{*}=z+k^{*}$, which satisfy the two constraints, 
Eqs. (\ref{eq:constr1}) and (\ref{eq:constr2}). 
The factor $q_1$ is the free monomer partition function while all $q_i$ factors 
($3 \leq i \leq z^{*}$) are partition functions of single anchored filaments of 
size $i$. 
We specify our model by choosing to describe our filaments as discrete wormlike 
chains with step length $d$ and persistence length $l_{\rm p}$, in contact with a 
hard wall when $L_{\rm c}>L$. 
This can be expressed as
\begin{align}
q_i(L)&=\lim_{k\rightarrow \infty}
\frac{1}{h^{3i}}\int dr^i \int dp^i \exp{(- \beta H_i)} 
\delta(x_1)\delta{(y_1)} \delta{(z_1)}  \delta(x_2-d) \delta{(y_2)} 
\delta{(z_2)},
\label{eq:qi}
\end{align}
where the $\delta$ functions impose the normal grafting of the filament in 
the plane located at $x=0$ and where $H_i$ is the Hamiltonian system of the 
filament of length $i$ with intramolecular interactions including a stiff 
bond potential with a force constant $k$ on which the limit is taken to fix 
the contour length. 
$H_i$ also contains external interactions with the obstacle wall located at 
$x=L$ but no other interactions with the grafting wall given the limited 
flexibility of the filaments. 
The explicit Hamiltonian $H_i$ reads
\begin{align}
H_i(\bm{r},\bm{p})&= 
\sum_{j=1}^i \frac{\bm{p}_j^2}{2 m} -(i-1) \epsilon_0' 
+ \sum_{j=1}^{i-1} \frac{k}{2}(d_j-d)^2
\nonumber\\
&+ \frac{\kappa}{d} \sum_{j=2}^{N-1}(1-\cos \theta_j)
+\frac{i-1}{2 \beta} \ln{\left[\frac{2 \pi}{\beta k d^2}\right]}
+U_i^{\rm ext},
\label{eq:hamil}
\end{align}
which starts with the kinetic energy term. 
The second term expresses the bonding energy corresponding to the energy 
released as heat when a new monomer attaches the filament and forms a new bond. 
There are $i-1$ bonds $\bm{d}_j=\bm{r}_{j+1}-\bm{r}_{j}$ of length $d$ as 
we will consider in the third term stiff harmonic springs with $d_j$ 
supposed to oscillate harmonically around $d$. 
The next term accounts for the bending energy where $\theta_j$ is the 
bending angle between bonds $\bm{d}_{j-1}$ and $\bm{d}_{j}$, while the 
bending modulus $\kappa$ fixes the persistence length $l_{\rm p}$ of the filament 
according to $\kappa=k_{\rm B}T l_{\rm p}$. 
The constant fifth term is needed to normalize the 
$\exp{\left[-\beta k(d_j-d)^2/2\right]}$ term which will appear in the 
filament canonical partition function (or in any phase space integral) in 
order to allow the (scalar) Gaussian bond length distribution around the mean 
$d$ to properly evolve towards the delta function in the 
$k \rightarrow \infty$ limit.
On this issue, it must be stressed that when taking the free filament 
Hamiltonian Eq.(\ref{eq:hamil}) without the external potential term, 
the integration of $\exp{(-\beta H)}$ over all Cartesian momenta and over 
all bond length variables $d_j$ followed by taking the infinitely stiff 
spring limit, leads to the usual discrete form of the wormlike chain 
effective configurational free energy
\begin{equation}
E_N[(\theta_j)_{j=2,(i-1)}]= 
-(i-1) \epsilon_0' +\frac{\kappa}{d} 
\sum_{j=2}^{i-1}(1-\cos \theta_j).
\label{eq:pot}
\end{equation}
The external potential term is specified by by a hard wall 
term $U^{\rm w}(L-x_j)$ so that
\begin{align}
U_i^{\rm ext}(L)=\sum_{j=z+1}^{i} U^{\rm w}(L-x_j),
\label{eq:ext}
\end{align}
where the lower index of the sum in Eq.(\ref{eq:ext}) refers to index 
$z(L)$ representing the largest filament size which does not interact 
with the obstacle. 
For the discrete WLC model, this index is given by
\begin{align}
z=1+I(L/d),
\label{eq:z}
\end{align}
where $I(x)$ denoted the integer part of a real value. 
For the finite stiffness filament model used in Molecular Dynamics type 
calculations \cite{Caby.12} or/and when soft repulsive walls with 
maximum range $R_{\rm c}$ are used, one should take 
\begin{align}
z=1+I(L^{\rm eff}/d),
\label{eq:zeff}
\end{align}
where $L^{\rm eff}=L-R_{\rm c}-(\beta k)^{-1/2}(L-R_{\rm c})/d$ in 
which the last term takes into account the bond fluctuations (of the order 
$1/\sqrt{\beta k}$) for a filament of contour length $\approx L-R_{\rm c}$.

\subsection{Filament partition function and the reactive canonical ensemble 
partition function of the bundle}

The ideal filaments bundle partition function given by Eq.(\ref{eq:rci}) 
requires all single monomer $q_1$ and single grafted filament $q_i$ 
partition functions. 
The free monomers are restricted in the pore volume $V=AL$ so that for 
the ideal solution,
\begin{equation}
q_1=\frac{V}{\Lambda^3},
\label{eq:q1}
\end{equation}
where the integration of the canonical partition function over momenta 
leads to the free monomer thermal de Broglie wavelength 
$\Lambda=\sqrt{\beta h^2/2 \pi m}$.

If we denote as $q_i^0$ the partition function of the same grafted 
filament system in absence of the obstacle wall, we can thus write
\begin{align}
q_i&=q_i^0 \;\;(3 \leq i \leq z), \nonumber \\
q_i&\equiv \alpha_i(L) q_i^0 \;\;(z< i \leq z^{*}), 
\label{eq:qq}
\end{align}
where the second relationship defines the wall factor 
$\alpha_i(L)$, which also depends on parameters $l_{\rm p},d$ and $T$. 
It is defined formally as
\begin{align}
\alpha_i(L)&=\frac{q_i(L)}{q_i^0}=
\langle \exp{(-\beta U_i^{\rm ext})}\rangle_{i0},
\label{eq:aa}
\end{align}
where $\langle\ldots\rangle_{i0}$ is a single filament canonical average 
with the Hamiltonian $H_i$ given by Eq. (\ref{eq:hamil}) and explicit 
grafting constraints (see Eq. (\ref{eq:qi})) but with the $U_i^{\rm ext}$ 
term turned off.

The equilibrium constant $K_i$ of the chemical reaction series in 
Eq. (\ref{eq:reaction}) in the ideal solution for grafted filaments 
with obstacle wall present, is given by \cite{b.Hill}
\begin{align}
K_{i}&=\frac{q_i}{q_{i-1} q_1/V}
=\frac{q_i}{q_{i-1}} \Lambda^3
=\frac{\alpha_i}{\alpha_{i-1}} 
\frac{q_i^0}{q_{i-1}^0}\Lambda^3 \equiv 
\frac{\alpha_i}{\alpha_{i-1}} K^0,
\label{eq:Ki}
\end{align}
where $\alpha_i=1$ for $i \leq z$ and $\alpha_i$ is given by 
Eq. (\ref{eq:aa}) for $i>z$, and where we have considered that the 
equilibrium constant for the chemical reaction in absence of obstacle, 
denoted as $K^0$, is independent of $i$. 
For our grafted filament model with stiff bonds defined by the 
Hamiltonian in Eq. (\ref{eq:hamil}), the equilibrium constant $K^0$ is 
indeed given by the $i$ independent expression
\begin{align}
K_0&=\frac{q_i^0}{q_{i-1}^0}\Lambda^3 
= 2\pi \exp{(\beta \epsilon_0')}\; 
\frac{d^4}{l_{\rm p}}\;\left[1-
\exp{\left(-2\frac{l_{\rm p}}{d}\right)}\right] F(w_0),
\label{eq:ratio}
\end{align} 
where the last factor, involving $w_0=\sqrt{\beta k d^2}$, 
is a correcting factor for bond flexibility with relative fluctuations 
$\sigma_d/d=w_0^{-1}$, 
\begin{align}
F(w_0)&= 
\left(\frac{w_0^2}{2 \pi}\right)^{1/2} 
\int_0^{\infty}du \;
\exp{\left[-\frac{w_0^2}{2} (u-1)^2\right]}\nonumber \\
&=\frac{\left[\frac{1+{\rm erf}[w_0]}{2}(1+w_0^2)
+ \frac{w_0}{\sqrt{2 \pi}} \exp{(-\frac{w_0^2}{2})}\right]}{w_0^2}.
\label{eq:fw0}
\end{align} 
As $F(w_0)$ satisfies the property 
$\lim_{k \rightarrow \infty}F(w_0)=1$, the equilibrium constant of the 
discrete WLC, hence the superscript dWLC, follows as
\begin{align}
K_0^{\rm dWLC}&= 2\pi \exp{(\beta \epsilon_0')}\; 
\frac{d^4}{l_{\rm p}}\;\left[1-\exp{(-2l_{\rm p}/d)}\right].
\label{eq:ratio}
\end{align}
Exploiting Eqs.(\ref{eq:q1}), (\ref{eq:qq}), (\ref{eq:aa}) and (\ref{eq:Ki}), 
we finally rewrite all single filaments partitions as
\begin{align}
q_{i}&= q_3^0 \left(\frac{K^0}{\Lambda^3}\right)^{i-3}\ 
(3 \leq i \leq z), \label{eq:qif1}\\
q_{i}(L)&= \alpha_i(L) q_3^0 
\left(\frac{K^0}{\Lambda^3}\right)^{i-3} \ (z < i \leq z^{*}).
\label{eq:qif2} 
\end{align}
Substituting in Eq. (\ref{eq:rci}) the individual filament partition 
functions by Eqs. (\ref{eq:qif1}) and (\ref{eq:qif2}), leads to
\begin{align}
Q^{\rm RC}&= N_f!\;q_3^{N_f} 
\left(\frac{K^0}{\Lambda^3}\right)^{(N_t-3 N_f)} \nonumber\\
& \times \sum_{N_1,N_3,..}
\frac{\alpha_{z+1}^{N_{z+1}}..
\alpha_{z+k}^{N_{z+k}}..
\alpha_{z+k^{*}}^{N_{z+k^{*}}}}{N_1! N_3!   
...N_z! N_{z+1}!..N_{z+k}!..} 
\left(\frac{K^0}{V}\right)^{-N_1}.
\label{Qrc}
\end{align}

\subsection{Filament size distribution}

The average number densities of free monomers and filaments in the 
thermodynamic limit can be estimated by searching for the largest term 
(constrained extremum) of the partition function $Q^{\rm RC}$ in 
Eq. (\ref{Qrc}), proceeding exactly like in the case of free 
filaments~\cite{b.Hill,Caby.12}. 
This requires solving the constrained global minimum
\begin{subequations}
\label{eq:deriv}
\begin{align}
&\frac{\partial }{\partial N_1} \ln 
\left[\frac{1}{N_1!}\left(\frac{K^0}{V}\right)^{-N_1}\right]-\lambda =0, \\
&\frac{\partial }{\partial N_i} 
\ln \left[\frac{1}{N_i!}\right]-i \lambda - \mu=0\quad \quad (i=3,z),\\
&\frac{\partial }{\partial N_{z+k}} 
\ln \left[\frac{\alpha_k^{N_{z+k}}}{N_{z+k}!}\right]- (z+k) \lambda - \mu
=0\quad\quad \ (k=1,k^{*}),
\end{align}
\end{subequations}
where $\lambda$ and $\mu$ are Lagrange multipliers related respectively to 
the constraints given by Eqs. (\ref{eq:constr1}) and (\ref{eq:constr2}). 
Use of the Stirling's approximation leads to the set of equations
\begin{align}
&\ln N_1 + \ln \left(\frac{K^0}{V}\right)+ \lambda =0, \\
&\ln N_i + i \lambda - \mu =0
\quad\quad\quad \quad (i=3,z),\\
&\ln (N_{z+k}/\alpha_{z+k}) + (z+k) \lambda - \mu=0\quad \quad (k=1,k^{*}).
\end{align}
In terms of the free monomer reduced number density 
$\hat{\rho}_1=\rho_1 K^0$, one gets
\begin{align}
\hat{\rho}_1 &= \exp{(-\lambda)}, \\
N_i &= \exp{(- [i \lambda + \mu])}= 
\hat{\rho}_1^{i} \exp{(- \mu)}
\quad\quad (i=3,z),
\label{eq:Ni}\\
N_{z+k} &= \alpha_{z+k} \exp{(- [(z+k) \lambda + \mu])}, \nonumber \\
 &=\alpha_k \hat{\rho}_1^{z+k} \exp{(- \mu)}
\quad \quad\quad\quad\quad\quad (k=1,k^{*}).
\label{eq:Nzk}
\end{align}
The combination of Eqs.~(\ref{eq:constr2}),~(\ref{eq:Ni}) and ~(\ref{eq:Nzk}) 
gives $\exp{(-\mu)}=N_f/D$ where $D$ and filament densities are finally given 
(with $L$ dependence made explicit) by
\begin{align}
P_{i}(\hat{\rho}_1,L)&= \frac{N_i}{N_f} 
= \frac{(\hat{\rho}_1)^{i}}{D} 
\quad \quad \quad \quad \quad\quad (3 \leq i \leq z(L)), 
\label{eq:popi}\\
P_{z+k}(\hat{\rho}_1,L)&= \frac{N_{z+k}}{N_f} 
= \alpha_{z+k}(L) \frac{(\hat{\rho}_1)^{z+k}}{D} 
\quad \quad \quad \quad (1 \leq k \leq k^{*}),
\label{eq:pop} \\
D(\hat{\rho}_1,L) &= 
\left[\sum_{i=3}^z (\hat{\rho}_1)^{i}\right] 
+ \sum_{k=1}^{k^{*}} \alpha_{z+k}(L) (\hat{\rho}_1)^{z+k}, \label{eq:ddd}
\end{align}
where $z$ is fixed by the condition Eqs. (\ref{eq:z}) or (\ref{eq:zeff}) 
and where $z^{*}$ was fixed by Eq. (\ref{eq:z*}). 
The reduced density $\hat{\rho}_1$ is itself the solution of an implicit 
equation obtained by substituting the filament densities Eqs. (\ref{eq:popi}) 
and (\ref{eq:pop}) in the constraint relationship Eq. (\ref{eq:constr1}), 
giving
\begin{equation}
\rho_t= \rho_1+\rho_f \langle i \rangle
=\rho_1+\rho_f \frac{M(\hat{\rho}_1,L)}{D(\hat{\rho}_1,L)},
\label{eq:rho1}
\end{equation}
where $\rho_t$, $\rho_f$ are respectively the total monomer and filament 
number densities and where $\langle i \rangle$ is the average length of the 
filaments in the bundle. 
In Eq. (\ref{eq:rho1}), $D$ is given by Eq. (\ref{eq:ddd}) and $M$ by
\begin{equation}
M(\hat{\rho}_1,L)=
\sum_{i=3}^{z} i (\hat{\rho}_1)^{i}
+\sum_{k=1}^{k^{*}} (z+k) \alpha_{z+k}(L)(\hat{\rho}_1)^{z+k}
\equiv \hat{\rho}_1 \frac{\partial D}{\partial \hat{\rho}_1}.
\label{eq:num}
\end{equation}

\subsection{Bundle pressure and individual filament force exerted on the opposite wall}

Equations (\ref{eq:popi}), (\ref{eq:pop}) and (\ref{eq:rho1}) provide 
the population densities, compatible with the constraints, which 
correspond to the largest term in the reactive canonical partition function 
$Q^{\rm RC}$ (\ref{Qrc}) and which can be identified, in the thermodynamic limit, 
as the equilibrium densities $\langle N_1\rangle/V$ or 
$\langle N_i \rangle/V$. 
Restricting the partition function to its maximum term 
$\tilde{Q}^{\rm RC}$, 
we can now obtain the partial derivatives in Eqs. (\ref{eq:pN}) and 
(\ref{eq:mu1}) from the Helmholtz free energy
\begin{align}
-\beta F=& \ln \tilde{Q}^{\rm RC}= C(N_f,T) 
+ N_t \ln{\left(\frac{K_0}{\Lambda^3}\right)}
+\sum_{k=1}^{k^{*}} 
\langle N_{z+k}\rangle \ln{\alpha_{z+k}} \nonumber \\
 & + N_1- N_1 
\ln{(\langle N_1\rangle K_0/V)} 
- \sum_{i=3}^{z^{*}} \langle N_{i}\rangle 
\ln{\langle N_i\rangle} + N_{f},
\end{align}
where $C(N_f,T)=\ln{\left[N_f!\;
(K_0/\Lambda^3)^{-3 N_f} q_3^{N_f}\right]}$, 
using chemical densities provided by Eqs.(\ref{eq:popi}), (\ref{eq:pop}) 
and (\ref{eq:rho1}). 
Substituting for the number of filaments at equilibrium, one gets 
\begin{align}
-\beta F&= C(N_f,T) + N_t \ln{\left(\frac{K_0}{\Lambda^3}\right)}
+ N_1 - N_1 \ln{\hat{\rho}_1} \nonumber \\
  & - N_f \langle i\rangle \ln{\hat{\rho}_1} 
+ N_f - N_f \ln{N_f} + N_f \ln{D}.
\label{eq:part1}
\end{align}
Using the conservation of the total number of monomers  
$N_t=N_1 + N_f \langle i\rangle$, see Eq. (\ref{eq:rho1}), 
the free energy Eq. (\ref{eq:part1}) can be simplified and expressed in 
terms of independent variables and in terms of the intermediate 
$\hat{\rho}_1$, given the lack of an explicit 
$\hat{\rho}_1(N_t,N_f,A,L,T)$ dependency,
\begin{align}  
\beta F = & C'(N_f,T) 
- \frac{A L}{K_0(T)}\hat{\rho}_1 
   + N_t \ln
\left(
\frac{\hat{\rho}_1 \Lambda^3(T)}{K_0(T)}\right) 
- N_f \ln{D(\hat{\rho}_1,L)},
\label{eq:part2}
\end{align}
where we have introduced $C'=-C-N_f+N_f \ln{N_f}$.

We first check the consistency of the final free energy expression by 
combining Eqs. (\ref{eq:mu1}) and (\ref{eq:part2}) to get the ideal 
solution expression of the chemical potential $\mu_1$, 
\begin{align}
\beta \mu_1 &= \left(\frac{\partial (\beta F)}{\partial N_t}\right) 
+ \left(\frac{\partial (\beta F) }{\partial \hat{\rho}_1}\right) 
\left(\frac{\partial \hat{\rho}_1}{\partial N_t}\right),\\
 & = \ln{(\rho_1\Lambda^3)}- \left(\frac{AL}{K_0}
- \frac{N_t}{\hat{\rho}_1}
+ \frac{N_f \langle i \rangle} {\hat{\rho}_1}\right) 
\left(\frac{\partial \hat{\rho}_1}{\partial N_t}\right),\\
 & = \ln{(\rho_1\Lambda^3)},
\label{eq:muid}
\end{align}
where the expected result for an ideal solution follows by successively 
exploiting Eqs.(\ref{eq:num}) and (\ref{eq:rho1}). 

Combining similarly Eqs. (\ref{eq:pN}) and (\ref{eq:part2}) , 
we get the pressure $p_N$ 
\begin{align}
\beta p_N&= -\left(\frac{\partial (\beta F/A)}{\partial L}\right) 
- \left(\frac{\partial (\beta F/A) }{\partial \hat{\rho}_1}\right) 
\left(\frac{\partial \hat{\rho}_1}{\partial L}\right),\\
 & =  \rho_1 + \sigma_f \left(\frac{\partial 
\ln{D}}{\partial L}\right)_{\hat{\rho}_1},
\label{eq:paint}
\end{align}
where $\sigma_f=N_f/A$ is the filament surface density.
Isolating the osmotic pressure $\Pi=p-p^{*}$ where $p^{*}=\rho_1 k_{\rm B}T$, 
one finally gets the osmotic normal force per filament 
$f_N(L,\hat{\rho}_1)$ in an ideal solution as
\begin{align}
\beta f_N&= \frac{\pi}{\sigma_f k_{\rm B}T}
= \left(\frac{\partial \ln{D}}{\partial L}\right),
 \label{eq:DL}\\
 &=\frac{\sum_{k=1}^{k^{*}} \frac{\partial \alpha_{z+k}(L)}{\partial L} 
(\hat{\rho}_1)^{z+k}}{D} 
= \sum_{k=1}^{k^{*}} \frac{\partial \ln{\alpha_{z+k}(L)}}{\partial L} 
P_{z+k}(\hat{\rho}_1,L),
 \nonumber \\
 &=\sum_{k=1}^{k^{*}} \beta \bm{f}_{z+k}(L) P_{z+k}(\hat{\rho}_1,L),
\label{eq:fn}
\end{align}
where we have introduced a fixed length filament mean force and associated 
mean force potential
\begin{align}
\bm{f}_{z+k}(L)&=
- \frac{\partial W_{z+k}(L)}{\partial L}, 
\label{eq:mpforce}\\
W_{z+k}(L)&=-k_{\rm B}T \ln{\alpha_{z+k}(L)}.
\label{eq:mp}
\end{align}
Equation (\ref{eq:fn}) gives the equilibrium force exerted on a living grafted
filament by a fixed planar wall located at a distance $L$ from the wall 
to which the filament is grafted. 
As expected, it is a weighted average of the force exerted by the wall on a 
fixed length grafted filament (an average over its internal degrees of freedom) where 
each filament size has an absolute probability $P_{z+k}(\hat{\rho}_1,L)$. 
Of course, only the filaments longer than $z$ contribute.

\subsection{Chemical thermodynamic conditions for the applicability of the 
above theory}

Apart for the ideal solution conditions which allow us to treat a 
chemical equilibrium of otherwise independent entities, we insisted from 
the beginning about the necessary condition $P_{z^{*}}\cong 0$, where $z^{*}$ 
is given by Eq. (\ref{eq:z*}). 
Indeed, especially in supercritical conditions where filaments tend to grow 
continuously, the filaments must be sufficiently rigid to resist a bending turn of 90 degrees with 
respect to their initial orientation at grafting, normal to the wall. One should avoid that, by exploiting a rare but 
possible bending thermal fluctuation, the filament could find the way to continue its polymerization along the obstacle 
wall with no further bending energy penalty. This can be imposed by requesting that
\begin{align}
\frac{P_{z^{*}}(L,\hat{\rho_1})}{P_z(\hat{\rho_1})} 
= \alpha_{z^{*}}(L) \hat{\rho_1}^{(z^{*}-z)} \ll 1,
\end{align}
which implies, using for the force at any compression ($L_{ci}-L$), the buckling force expression for a filament of size 
$i$ and contour length $L_{{\rm c}i}$, namely \cite{frey.06}
$\beta f_{bi}= (\pi^2/4)  (l_{\rm p}/L_{{\rm c}i}^2)$, the inequalities
\begin{align}
- \beta W_{z^{*}}(L)+ (z^{*}-z) 
\ln{\hat{\rho}_1}  &< 0,  \nonumber \\
- \beta f_{bz^{*}}(\frac{\pi}{2}-1) L 
+ (\frac{\pi}{2}-1) \frac{L}{d} \ln{\hat{\rho}_1} &<0, 
\end{align}
and so finally, the condition on the reduced density
\begin{align}
\ln{\hat{\rho}_1} < \frac{l_{\rm p} d}{L^2}\equiv 
\ln{\hat{\rho}_{1b}},
\label{eq:rho1lim}
\end{align}
where $\hat{\rho}_{1b}$ is the upper limit of the reduced free monomer density.

\section{Monte-Carlo determination of the wall factors and 
illustrative applications
\label{sec:Monte-Carlo}}

\subsection{Monte-Carlo method}

We consider a filament modelled by the Hamiltonian $H_i$ given in 
Eq. (\ref{eq:hamil}) subject to grafting conditions at the wall in the 
plane at $x=0$ as formulated in Eq. (\ref{eq:qi}). 
Eq. (\ref{eq:ext}) provides the external potential due to the sum of 
individual interactions $U^{\rm w}(L-x)$ between any monomer in the range of 
interaction ($L-r_c<x<L$) and the obstacle wall located at $x=L$. 
We consider generally a large but finite stretching force constant 
$k$ but we will also be interested in the discrete WLC model obtained in 
the $k \rightarrow \infty$ limit. 
The conditions Eqs. (\ref{eq:z}) and (\ref{eq:zeff}) define the portion 
of the filament which does not interact with the wall directly.
The wall factor for a grafted filament of size $i>z$ is given formally by 
Eq.(\ref{eq:aa}) which can be rewritten as
\begin{align}
\alpha_i(L)&=
\left\langle\exp{(-\sum_{k=z+1}^{i} \beta U^{\rm w}(L-x_k))}\right\rangle_{i0}.
\label{eq:aa1}
\end{align}
Given the additivity of the various contributions to the Hamiltonian 
$H_i$ of a grafted filament in absence of obstacle wall and given the various 
simplifications between similar integrals in the numerator and the denominator 
in the average Eq. (\ref{eq:aa1}), one can rewrite the average Eq. (\ref{eq:aa1}) over an ensemble 
associated to a single filament of size $i$ as another average over a single filament of 
size $z$, in which the quantity to be averaged still contains an explicit 
integration of the remaining portion of the filament between $z+1$ and 
$i$, which must still be performed for each microscopic configuration of the main 
filament portion of size $z$. 
Using spherical coordinates $t_j,\eta_j=\cos \theta_j,\phi_j$ in 
successive local Cartesian coordinate systems for the extra bonds of 
index $j=z,z+1,\ldots,i-1$ (bond $j$ connects monomer $j$ and $j+1$), 
one gets
\begin{align}
\alpha_i(L,T)&=
\left(\frac{1}{F(w_0)}\right)^{i-z}  
\left\langle \prod_{j=z}^{i-1} \int d\bm{W}_j 
\exp{\left[-\beta U^{w}(L-x_{j+1})\right]} 
\right\rangle_{z} ,
\label{eq:alpha} 
\end{align}
where the average $\langle\ldots\rangle_z$ indicates an average over the 
configuration space of a grafted filament of size $z$ which does not 
interact with any wall, where $x_{j+1}$ is the $x$ coordinate of the monomer 
of index $j+1$ depending on the coordinates of the primary grafted chain of 
size $z$ and of the additional sampled coordinates of the bonds 
$z,z+1,\ldots,z+j-1$ of the extra piece of the filament. 
$U^{w}(L-x)$ is the wall potential acting on a monomer located at 
$x<L$. 
Finally, the integration $d\bm{W}_j$ over the spherical coordinates 
$t_j,\eta_j=\cos \theta_j,\phi_j$ is given by
\begin{align}
\int d\bm{W_j}& \equiv \int_0^{\infty}du_j\;u_j^2 \int_{-1}^{+1}d\eta_j 
\int_0^{2\pi} d\phi_j P_u(u_j) P_{\eta}(\eta_j) P_{\phi}(\phi_j),
\label{eq:not1}
\end{align}
in terms, for any bond, of normalized distribution functions of 
$u=t/d$, $\eta$ and $\phi$ 
\begin{align}
P_u&=\frac{1}{\sqrt{2\pi (\beta k d^2)^{-1}}}
\exp{\left(-\frac{(u-1)^2}{2 (\beta k d^2)^{-1}}\right)}, \label{eq:Pu}\\
P_{\eta}&=\frac{\exp{(-l_{\rm p}(1-\eta)/d)}}
{ (d/l_{\rm p}) [1-\exp{(-2 l_{\rm p}/d)}]},\label{eq:Peta}\\
P_{\phi}&=\frac{1}{2\pi}.
\label{eq:Pphi}
\end{align}
Note that this writing (with $P_u$ improperly normalized as a Gaussian 
distribution which would sample $u$ in the $[-\infty,+\infty]$ range), 
it is possible to replace $P_u$ by the $\delta(u-1)$ function as 
$k \rightarrow \infty$. 

Alternatively, using $P'_u$ properly normalized on the 
$[0, +\infty]$ range, we get 
\begin{align}
\alpha_i(L,T)&=\left(\frac{V(w_0)}{F(w_0)}\right)^{i-z}  
\left\langle \prod_{j=z}^{i-1} \int d\bm{W'}_j 
\exp{\left[-\beta U^{w}(L-x_{j+1})\right]} \right\rangle_{z}, 
\label{eq:alphab} 
\end{align}
where 
\begin{align}
\int d\bm{W'_j}& \equiv \int_0^{\infty}du_j\;u_j^2 
\int_{-1}^{+1}d\eta_j \int_0^{2\pi} d\phi_j 
P'_u(u_j) P_{\eta}(\eta_j) P_{\phi}(\phi_j),
\label{eq:not1}
\end{align}
with
\begin{align}
V(w_0)&= \frac{\int_0^{\infty}
\exp{(-\frac{(u-1)^2}{2 \sigma_d^2})}}
{\int_{-\infty}^{\infty}
\exp{(-\frac{(u-1)^2}{2 \sigma_d^2})}}=\frac{1+{\rm erf}[w_0]}{2},\\
P'_u&=\frac{\exp{(-\frac{(u-1)^2}{2 \sigma_d^2})}}
{\int_{0}^{\infty}\exp{(-\frac{(u-1)^2}{2 \sigma_d^2})}}=P_u/V(w_0).
\label{eq:pprim}
\end{align}
The wall factor $\alpha_i$ in Eq. (\ref{eq:alphab}) can be computed as 
any average $\langle\ldots\rangle_z$ by using a standard Metropolis 
Monte-Carlo sampling the configuration variables of a grafted filament of 
size $z$. 
For each microscopic configuration of the primary filament, a simple 
Monte-Carlo procedure is used to sample the $3^{i-z}$ additional variables  
($u_{j},\eta_{j},\phi_{j}$) in distributions $P'_u$, $P_{\eta}$ and $P_{\phi}$ 
and hence solve numerically the multiple integral, what can be formulated
\begin{align}
\alpha_i(L)= \left(\frac{V(w_0)}{F(w_0)}\right)^{i-z}
 \left\langle \lim_{M\rightarrow\infty}
\frac{1}{M}\sum_{m=1}^M \left[\prod_{j=z}^{i-1} 
\left(u_{jm}^2 \exp{\left[-\beta U^{w}\left((x_{j+1})_m-L
\right)\right]}\right)\right] \right\rangle_z, 
\label{eq:genat}
\end{align}
where index $m$ denotes a particular sampling of $3(i-z)$ variables, 
where $(x_{j+1})_m$ is the $x$ coordinate of monomer $j+1$, 
a function of the primary filament configuration variables and also function 
of all variables obtained in the $m^{\rm th}$ sampling for all intermediate 
bonds $z,\ldots,z+j-1$.

In the specific case of a infinitely stiff discrete WLC interacting with a 
hard wall for which $\exp{(- \beta U^{\rm w}(L-x))}=\Theta(L-x)$ where 
$\Theta(y)$ is the Heaviside function, Eq. (\ref{eq:genat}) simplifies to
\begin{equation}
\alpha_i(L)= \lim_{M\rightarrow\infty} 
\left\langle \frac{1}{M} \sum_{m=1}^M \prod_{j=z}^{i-1} 
(\Theta(L-(x_{j+1})_m))\right\rangle_z.
\label{eq:wlchwt}
\end{equation}
Use of Eqs. (\ref{eq:genat}) and (\ref{eq:wlchwt}) require the sampling of 
$i-z$ additional bond variables in distributions Eqs. (\ref{eq:Peta})
and (\ref{eq:Pphi}). 
Each pair of  $\eta,\phi$ variables can be obtained from two random 
numbers $\xi_1$ and $\xi_2$ sampled in a uniform distribution $[0,1]$ 
according to
\begin{align}
\eta&=1+\frac{d}{l_{\rm p}} 
\ln{\left[\xi_1+(1-\xi_1)\exp\left(-\frac{2 l_{\rm p}}{d}\right)\right]},\\
\phi&=2 \pi \xi_2.
\end{align}
For sampling the $u$ variable in Eq. (\ref{eq:pprim}) for flexible bonds 
appearing in Eq. (\ref{eq:genat}), one samples a random number $\xi_3$ in 
a normal Gaussian distribution with zero mean and the reduced bond length 
follows as
\begin{equation}
u=\xi_3 \frac{\sigma_d}{d} +1=\xi_3 (w_0)^{-1}+1.
\end{equation}
If $u\geq0$, $u$ is accepted as a sampled reduced bond length while a 
value $u<0$ is simply rejected and a new $\xi_3$ sampling is performed 
(until a positive $u$ value is obtained).

To conclude this section, let us point out that the determination of wall 
factors by Monte-Carlo can be performed in various other ways. 
The above procedure was found to be rather efficient as, using a single MC 
Metropolis sampling for any given $z$ size primary filament, the 
$\alpha_{z+k}(L)$ can be simultaneously estimated for $k=1,k_{\rm max}$ 
where $k_{\rm max} \cong 5$ and for $L$ positions in the range going from 
$L_{\rm min}=(z-1)d$ up to a $L_{\rm max}=zd$. 
We systematically performed four similar but independent runs (each of 
$2 \times 10^6 \ (z-2)^2$ attempted steps) to use the dispersion of results 
to estimate the statistical errors. 
For the wall position variable, we used a discretization step 
$\Delta L=0.01d$ in order to estimate numerically the derivatives 
$\partial \alpha/\partial L$ required to estimate the fixed length 
filament wall forces, Eqs.(\ref{eq:mpforce}) and (\ref{eq:mp}).

\subsection{Illustrative results}

\subsubsection{The wall factors}

\begin{figure}
\begin{center}
\includegraphics[scale=0.4]{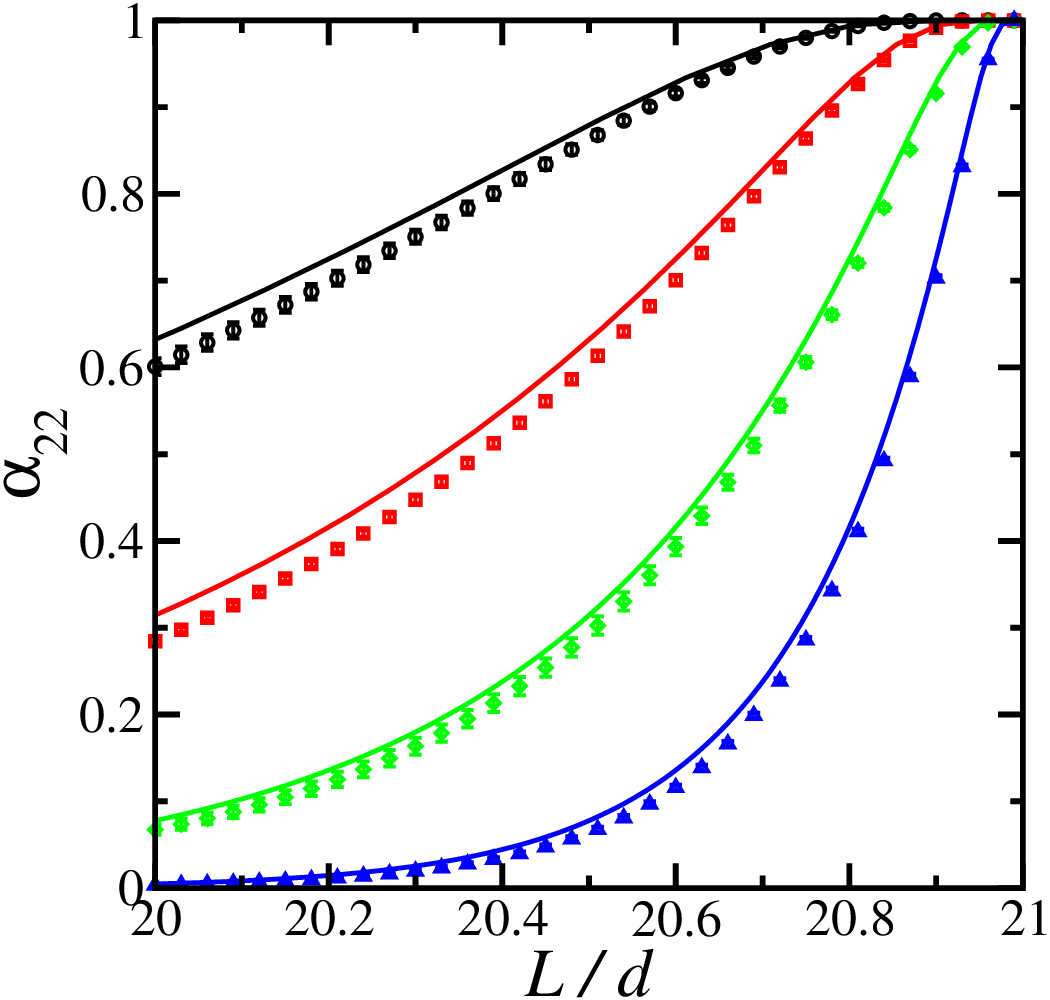}
\caption{The dependence of the wall-factor $\alpha_{22}(L)$ on position of 
the wall for the discrete WLC model, for four different $l_{\rm p}/d$ values, 
namely $1000$ (blue triangles), $500$ (green lozenges), $250$ (red squares) 
and $125$ (black circles). 
Statistical errors on $\alpha_{22}$ data are indicated by $\pm 1 \sigma$ 
error bars. 
The continuous curves are plots of the unique universal curve for a similarly 
grafted continuous WLC hitting a hard wall \cite{frey.06}, 
namely $\alpha_{22}(L,l_{\rm p})=\tilde{Z}(\tilde{\eta})$. 
Here, the definition of the scaling variable given by Eq. (\ref{eq:eta}) is 
exploited along with $L_{\parallel}=3.5d, 1.8d, 0.88d, 0.44d$ for the four 
$l_{\rm p}$'s in ascending order. 
Note that this continuous WLC theoretical curve is valid for the three 
largest $l_{\rm p}$'s but is supposed to be only approximately valid for 
$l_{\rm p}/d=125$ as the criterium for validity of the universal regime is that the stiffness parameter 
satisfies $L_{\rm c}/l_{\rm p}<0.1$ \cite{frey.06}.}
\label{fig:bb1}
\end{center}
\end{figure}

\begin{figure}[h!t]
\begin{center}
\includegraphics[scale=0.4]{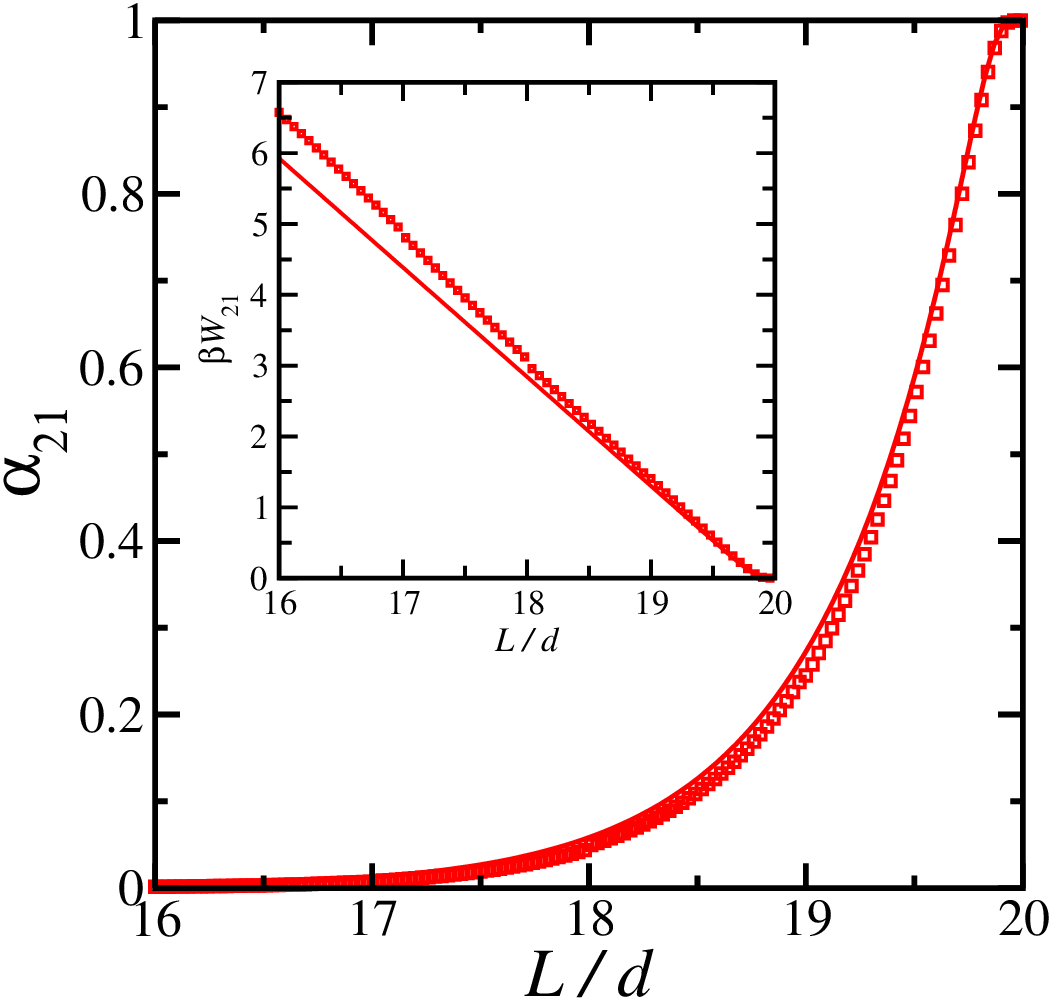}
\caption{The dependence of the wall factor $\alpha_{21}(L)$ of a discrete 
WLC with persistence length $l_{\rm p}/d=250$ over a range of wall positions 
$L$ in the range $L/d=16-20$. 
The Monte-Carlo data are shown by symbols (note that statistical error 
bars on $\alpha_{21}$ are smaller than the size of the symbols) while the 
continuous curve reproduces the universal curve 
$\alpha_{21}(L,l_{\rm p})=\tilde{Z}(\tilde{\eta})$ \cite{frey.06}, where the 
scaling variable $\tilde{\eta}$ is given by Eq. (\ref{eq:eta}) with $L_{\parallel}=1.6d$. 
In inset, the corresponding potential of mean force 
$\beta W_{21}=-\ln{\alpha_{21}}$ is again compared 
(symbols for MC data) to the continuous WLC universal curve 
$\beta \tilde{F}_{\parallel}(\tilde{\eta})$ \cite{frey.06}, 
adapted to the present case.}
\label{fig:bb2}
\end{center}
\end{figure}

In Figs. \ref{fig:bb1} and \ref{fig:bb2}, we illustrate the behaviour of 
$\alpha_i(L)$ defined by Eq. (\ref{eq:aa}) for a grafted discrete WLC 
hitting a hard wall, exploiting MC simulations data obtained via Eq.(\ref{eq:wlchwt}). 
We first show the results for a filament of size $i=21$ studied at moderate 
filament absolute compression, for various values of the persistence length. 
In all cases, the function reaches unity when the wall position $L$ reaches 
from below the contour length of the originally compressed wormlike chain 
filament, but in a way which gets steeper as $l_{\rm p}$ increases. 
Our MC results can be compared to the scaling properties of the continuous 
WLC model for a grafted filament of contour length $L_{\rm c}$ and persistence 
length $l_{\rm p}$ hitting a hard wall located at $L$ \cite{frey.06}. 
The compression variable must be rescaled by a length 
$L_{\parallel}$, according to
\begin{align}
\tilde{\eta}&=\frac{(L_{\rm c}-L)}{L_{\parallel}}, \label{eq:eta} \\
L_{\parallel}&=\frac{L_{c}^2}{l_{\rm p}}.
\label{eq:Lpar}
\end{align}
Adapting the general theoretical law to the specific cases shown in 
Figs. \ref{fig:bb1} and \ref{fig:bb2}, we get close but distinct results 
with respect to our MC data dealing with a discrete chain. 
The MC results show systematically (also for other cases explored but not 
shown here) that the discrete chain (with respect to the continuous filament) 
requires more energy to be compressed by a similar absolute amount $L_c-L$.
In Fig. \ref{fig:bb2}, the behaviour of the wall coefficient $\alpha_{21}(L)$ 
is shown overs a large $L$ window which was obtained by assembling the results 
from four independent runs covering a different $L$ window of size $d$. 
The comparison with the scaling law \cite{frey.06} shows again similarities 
and a somewhat stiffer behaviour of the discrete WLC (with step size $d$) 
with respect to its continuous limit, for identical 
$l_{\rm p}, L_{\rm c}$ and  $L$ parameters values.


\subsubsection{Distributions of living filament sizes}

\begin{figure}[h!t]
\begin{center}
\includegraphics[scale=0.4]{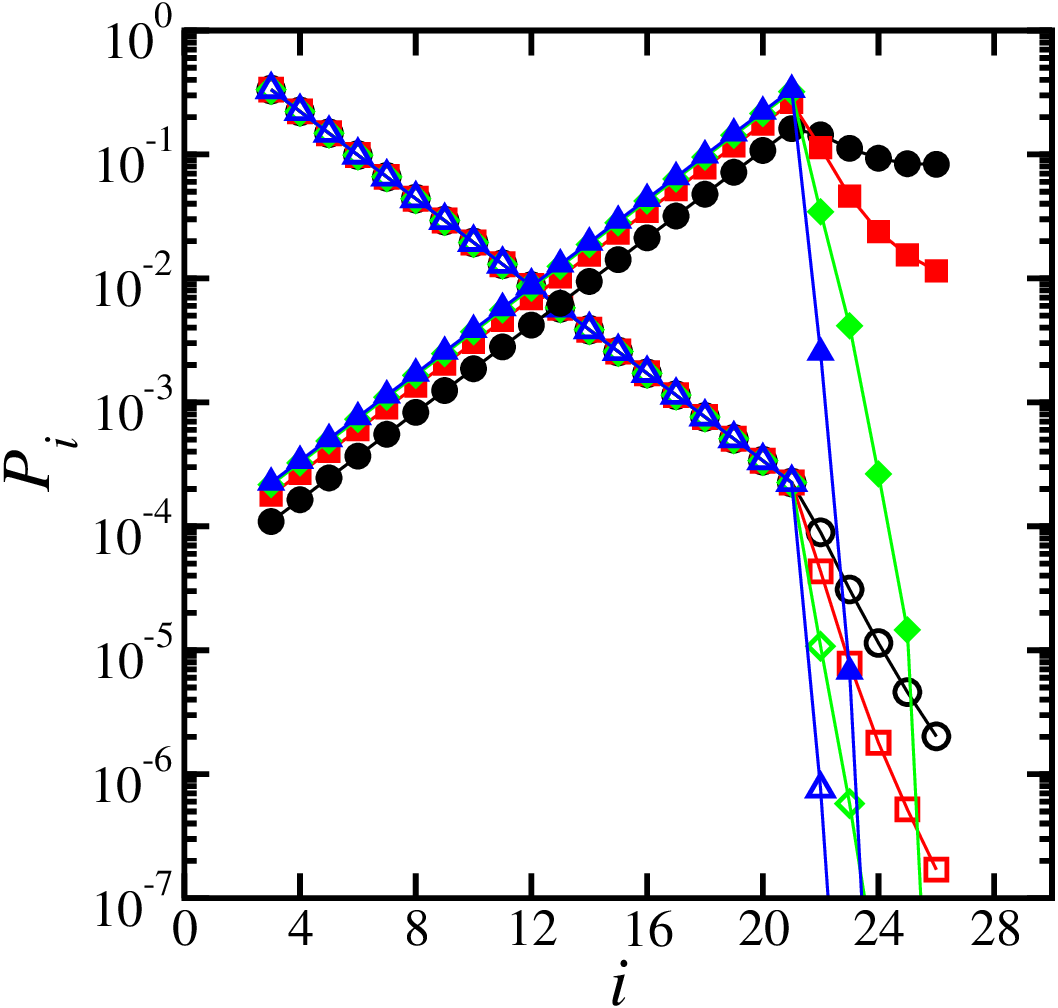}
\caption{Distributions of discrete WLC filament sizes, normalized in the 
$i=[3,z+k_{max}]$ window ($k_{max}=5$), for a bundle of living filaments 
grafted at the left wall and confined by a hard obstacle wall with gap 
size $L=20d$ at subcritical $\hat{\rho}_1=0.67$ (open symbols) and at 
supercritical $\hat{\rho}_1=1.5$ (filled symbols) conditions for four 
different $l_{\rm p}/d$ values, namely $1000$ (blue triangles), $500$ 
(green lozenges), $250$ (red squares) and $125$ (black circles).}
\label{fig:aa}
\end{center}
\end{figure}

\begin{figure}[h!t]
\begin{center}
\includegraphics[scale=0.4]{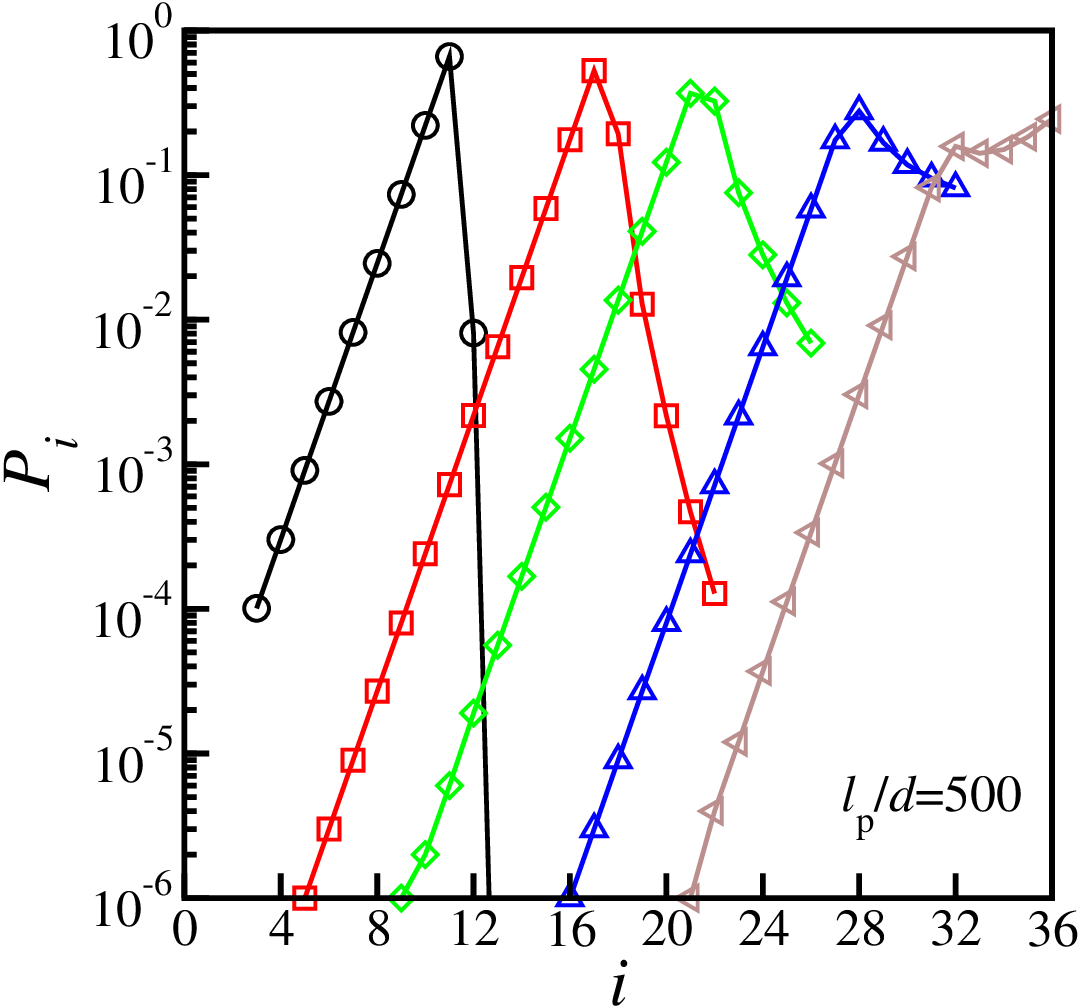}
\caption{Distributions of WLC filament sizes, normalized in the 
$i=[3,z+k_{max}]$ window ($k_{max}=5$), for filaments with persistence 
length $l_{\rm p}/d=500$ at $\hat{\rho}_1=3.00$ for various gap widths 
$L/d$ with associated $z(L)$ index value, namely from left to right, 
$L/d=10.5$ (circles), $16.5$ (squares), $20.5$ (lozenges), $26.5$ 
(dark triangles) and $30.5$ (light triangles).}
\label{fig:bb}
\end{center}
\end{figure}

Filament size distributions given in Eqs. (\ref{eq:popi}), (\ref{eq:pop}) 
and (\ref{eq:ddd}) are illustrated in Figs. \ref{fig:aa} and \ref{fig:bb} 
for a living, grafted (discrete) WLC hitting a hard wall, when exploiting the 
Monte-Carlo determined first five wall factors $\alpha_{z+k}(L)$  beyond the 
index $z$ fixed by the gap size $L/d$ (cfr Figs.\ref{fig:bb1} and 
\ref{fig:bb2}). 
In Fig. \ref{fig:aa}, the gap size $L/d=20$ and hence the $z=21$ value are 
fixed and the distributions are compared for various $l_{\rm p}$'s while in 
Fig. \ref{fig:bb}, the distributions are shown for various gap sizes $L/d$, 
for a fixed persistence length $l_{\rm p}/d=500$.
At the subcritical monomer density, one recognises in Fig. \ref{fig:aa} an 
exponential decay of short filaments densities followed by an even faster decay 
for the rare length fluctuations where filament sizes are able to hit the 
obstacle with their free ends. 
In the supercritical case, the wall interrupts the rising exponential 
distribution of short filaments densities but, for the same given pair of 
($L,\hat{\rho}_1$) parameters, the $l_{\rm p}$ value, and hence the bending 
susceptibility, is seen to strongly influence the decay of the distribution 
beyond the size $i=z$.
In Fig.(\ref{fig:bb}), we fix the persistence length to $l_{\rm p}=500d$ and fix 
the reduced free monomer density $\hat{\rho}_1$ to a rather large value and 
we compare the distributions for different hard wall gap sizes. 
We see that the convergence improves, as expected, when $L$ gets smaller for 
fixed $l_{\rm p},\hat{\rho}_1$ values.

The convergence of the large filament size populations in supercritical 
conditions should always be controlled by the criterium provided in  
Eq. (\ref{eq:rho1lim}). 
Fast convergence towards zero for populations of filaments hitting the 
wall and in particular the requirement $P(z^{*}) \cong 0$, where $z^{*}$ is 
defined by Eq. (\ref{eq:z*}), requires 
$\hat{\rho}_1 < \hat{\rho}_{1b}=\exp{\left(l_{\rm p} d/L^2\right)}$. 
For data in Fig. \ref{fig:aa} for which $z^{*}=31$ and $\hat{\rho}_1=1.50$, 
the inequality is verified for $l_{\rm p}/d=1000, 500$ and $250$ 
($\hat{\rho}_{1b}=12.2, 3.5$ and $1.9$ respectively) but the criterium is not 
verified for $l_{\rm p}/d=125$ ($\hat{\rho}_{1b}=1.4$). 
This is coherent with the observed trends obtained numerically for the 
different $l_{\rm p}$'s for filaments sizes $i=22$ up to $i=26 <z^{*}$, the lowest 
value of $l_{\rm p}$ suggesting a finite minimum in the distribution of filament 
lengths around a size $z+1<i<z^{*}$.  
In Fig. \ref{fig:bb} where all data refer to $\hat{\rho}_1=3$, the same 
criterium is marginally verified for the gap sizes $L/d=20.5$ with 
$\hat{\rho}_{1b}=3.3$ but not for the next gap size of 
$L/d=26.5$ ($z^{*}=42$) for which $\hat{\rho}_{1b}=2.0$. 
The poor convergence towards zero observed for the two wider gap sizes is 
again in agreement with the observation that $\hat{\rho}_1 > \hat{\rho}_{1b}$.

\subsubsection{Equilibrium force exerted by a living filament on a fixed wall}

\begin{figure}
\begin{center}
\includegraphics[scale=0.4]{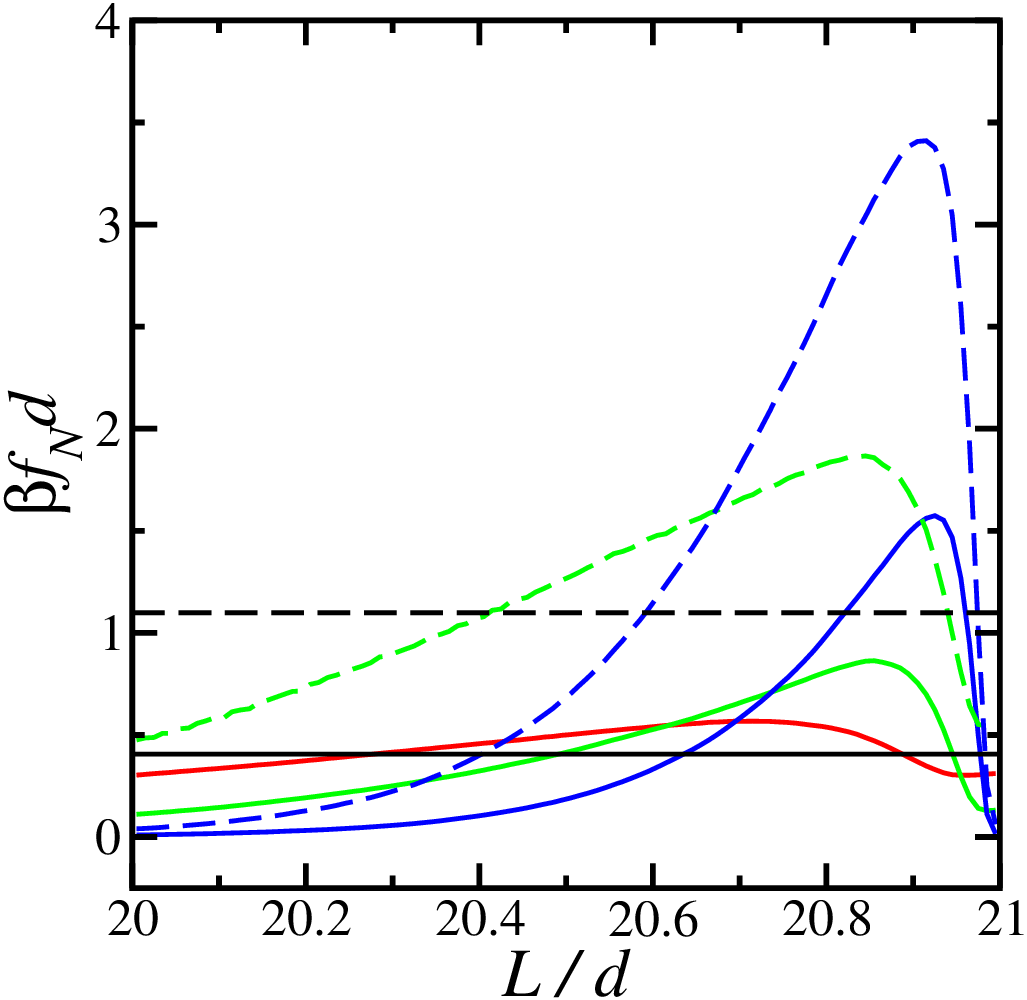}
\caption{For a right wall position between $L/d=20$ and $L/d=21$, evolution of 
the local reduced force $\beta f_N d$ exerted by the right hard wall on one 
WLC filament starting at the left wall with first monomer at $x=x_{wl}=0$. 
The continuous (dashed) curves correspond to free monomer reduced number 
densities $\hat{\rho}_{1}=1.5$ ($\hat{\rho}_{1}=3.0$) and colours indicate 
$l_{\rm p}$ values ($l_{\rm p}/d=250$ (red), $l_{\rm p}/d=500$ (green), $l_{\rm p}/d=1000$ (blue)). 
The case $l_{\rm p}/d=250$, $\hat{\rho}_{1}=3.0$ is not shown as 
$\hat{\rho}_1>\hat{\rho}_{1b}$. 
The black horizontal lines indicate levels of $\ln{\hat{\rho}_1}$, 
the continuous and dashed curves referring to $\hat{\rho}_{1}=1.5$ and 
$\hat{\rho}_{1}=3.0$ respectively.}
\label{fig:locfor}
\end{center}
\end{figure}

The normal force Eq. (\ref{eq:fn}) for a grafted filament hitting a wall 
can be computed for any value of the reduced free monomer density 
$\hat{\rho}_1<\hat{\rho}_{1b}$ provided the relevant series of wall factors 
$\alpha_{z+k}(L)$ and their derivatives with respect to $L$ are known. 
In Fig. \ref{fig:locfor}, we report some results for the discrete WLC model. 
The variations of the equilibrium force with the gap size $L$ over length 
scales below the monomer size $d$ are seen to be enhanced as the persistence 
length increases. 
The origin of these variations can be easily interpreted for rather stiff 
living filaments. 
It is known \cite{frey.06} that the compressive force on a grafted filament of 
fixed length $i$  increases from zero at $L=(i-1)d$ towards the limiting 
buckling force 
$f_c=(\pi^2/4) (k_{\rm B}T/L_{\parallel})$ where 
$L_{\parallel}$ is given by Eq. (\ref{eq:Lpar}), when the rescaled compression 
$\tilde{\eta}$ defined by Eq.(\ref{eq:eta}) is $\approx 0.4$. 
The behaviour of the equilibrium force in the range $L/d=20 \rightarrow 21$ 
for $l_{\rm p}/d=1000$ is essentially due to the action of filaments of length 
$i=22$ having a characteristic compression length 
$L_{\parallel}=21^2/1000=0.44$. 
Indeed there are no force contribution from shorter filaments and the 
probability to have longer filaments is marginal as it requires an absolute 
compression larger than $d$, hence a reversible work 
$\approx d f_c/(k_{\rm B}T)>5$. 
The observed increase of $f_N(L)$ with $L$ simply reflects the monotonic 
growth of the $P_{22}(L)$ probability over the whole interval which is 
multiplied by a constant $f_c$ until 
$(21-L/d) \approx 0.4 L_{\parallel} \approx 0.2$ where the force starts to 
decrease to zero as $L/d$ further increases to $21$.

Comparing forces for the same $l_{\rm p}$ and different reduced free monomer 
concentrations $\hat{\rho}_1$, we observe a systematic increase which can be 
quantified by the averaged force over the $d$ interval. 
Using Eq. (\ref{eq:DL}), the reduced averaged force turns out to be 
$D(L+d,\hat{\rho}_1)/ D(L,\hat{\rho}_1)\approx \ln{\hat{\rho}_1}$ 
as it can be qualitatively understood when noting that the wall factors 
appearing in $D(L+d,\hat{\rho}_1)$ and $D(L,\hat{\rho}_1)$ are in fact linked 
together by the approximate connection $\alpha_i(L+d) \approx \alpha_{i-1}(L)$. 
This result will be analyzed quantitatively and discussed in a biological 
context for the actin case in a separate publication \cite{X.13}.

We conclude these illustrations by the equilibrium force predicted for the 
grafted filament model that we used in direct simulations (using the Hybrid-Molecular Dynamics 
method) of a grafted bundle of interacting filaments, in chemical equilibrium with explicit free monomers, 
hitting the opposite wall \cite{RR.13}.  
The model (fully detailed in ref.\cite{Caby.12}) is just a sum of single 
filament Hamiltonian terms given in Eq. (\ref{eq:hamil}) and free monomers 
Hamiltonian terms (point-mass particles) to which excluded volume 
intermolecular forces are added. 
Purely repulsive Weeks-Chandler-Andersen (WCA) interactions (with LJ parameters 
$\sigma=0.891 d$ and $\epsilon/k_{\rm B}T=3$) operate between any pair of monomers 
belonging to different entities (entities are either free monomers or 
filaments). 
The monomer wall-interaction is taken as
\begin{equation}
U^{\rm w}(x)=\frac{3 \sqrt{3}}{2} 
\epsilon_{\rm w} \left[(\frac{\sigma_{\rm w}}{x})^9 
-(\frac{\sigma_{\rm w}}{x})^3 \right]
+\epsilon_{\rm w}
\end{equation}
where $x$ is the wall-monomer distance, where $\sigma_{\rm w}=d$ and 
$\epsilon_{\rm w}/k_{\rm B}T=0.1$. 
A cut-off is applied at potential minimum for $x_c=1.200936d$. 
In the simulations, the gap size was fixed to $L/d=16$ and the filament was 
modelled as follows: we adopt stiff bonds with a large but finite force 
constant $k=400 k_{\rm B}T/d^2$ giving rise to bond length fluctuations of the 
order $\sigma_d/d=0.05$ and we chose a persistence length of 
$l_{\rm p}=250d$ for the filaments.

\begin{figure}[h!t]
\begin{center}\includegraphics[scale=0.4]{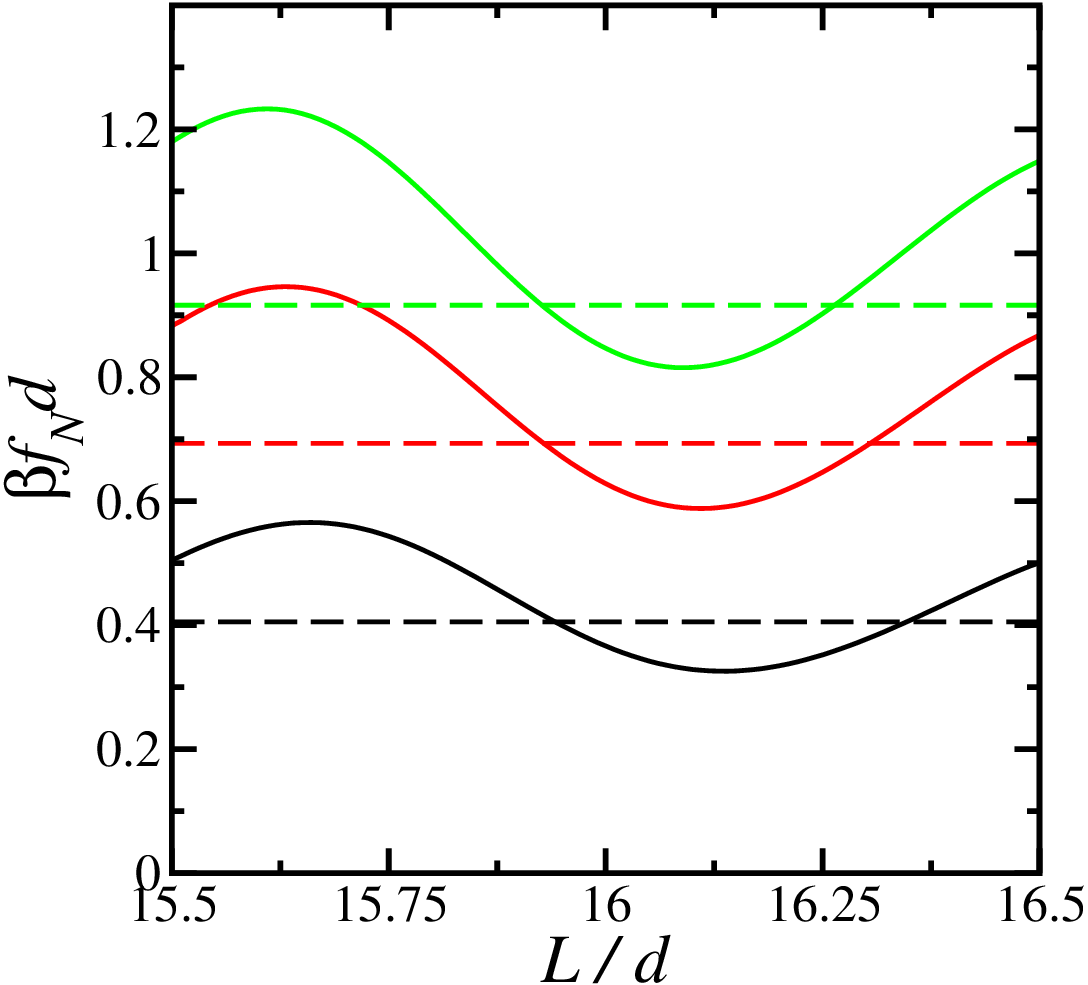}
\caption{Local reduced force $\beta f_N d$ for a right wall 
position varying between $L/d=15.5$ and $L/d=16.5$. 
The three continuous curves (from bottom upwards) correspond to free monomer reduced 
number densities $\hat{\rho}_{1}=1.5$, $\hat{\rho}_{1}=2.0$ and 
$\hat{\rho}_{1}=2.5$. Horizontal lines of specific colour indicate levels of 
the corresponding $\ln{\hat{\rho}_1}$ values, suggesting that the average 
force over the $d$ interval is somewhat larger in the present case.}
\label{fig:cc1}
\end{center}
\end{figure}

We have computed using Eqs. (\ref{eq:genat}) and (\ref{eq:zeff}) the first 
five wall factors $\alpha_{16}-\alpha_{20}$ for the filament and the 
filament-wall interaction associated to the above MD model and looked at the 
equilibrium normal force $f_N(L,\hat{\rho}_1)$ predicted in an ideal filament 
bundle solution where the present independent filament approach is relevant. 
The normal equilibrium force is shown in Fig. \ref{fig:cc1} in the range 
$15.5<L/d<16.5$ for three values of $\hat{\rho}_{1}$ below the limit 
$\hat{\rho}_{1b}=2.8$. 
We observe again large variations and an average reduced force of the 
order of $\ln{\hat{\rho}_1}$.

\section{Conclusions and perspectives \label{sec:concl}}

We have established the reactive canonical ensemble partition function 
corresponding to a set of independent living filaments undergoing 
(de)polymerizing reactions with free monomers from the bath, for a 
confined bundle system. 
Filaments are modelled as discrete wormlike chains grafted normally to one 
plane and confined by a second obstacle plane over a gap distance much 
shorter than the persistence length of the filaments. 
From this ideal system partition function, we have derived the distribution of 
filament sizes and the associated average force exerted by the living filaments 
on the opposite wall in supercritical conditions. 
We find strong oscillations of this equilibrium force with $L$ varying, while 
any $L$ averaged force over the single monomer size increment $d$ turns out 
to be of the order of the expected result, Eq. (\ref{eq:force}). 
A more detailed analysis of the flexibility and the gap size influences of this 
averaged force will be discussed elsewhere \cite{X.13}.
We have also provided a Metropolis Monte-Carlo procedure to compute the 
wall factors for various filaments/wall models. 
For the discrete wormlike chain hitting a hard wall, the results can be 
compared with the predictions of the continuous wormlike chain model for the 
same grafted filament/hard wall system \cite{frey.06}. 
The wall factors for the two models differ by a few percents, a result of the influence of the 
additional finite length scale $d$ which seems to lead to a slightly less 
flexible model than the classical continuous WLC for an identical persistence 
length. 
This work on ideal bundles at equilibrium could be extended towards the 
analysis of the non equilibrium force-velocity relationship for a set of 
independent filaments pushing on a mobile piston against an external load. 
A possible route is to consider as stochastic variables the individual sizes 
of the different filaments and the one dimensional position of the piston using 
to describe the filament compression the potential of mean force 
$W_i(L;l_{\rm p},d)=-k_{\rm B}\ln{\alpha_i(L;l_{\rm p},d)}$ which takes into 
account the compression of the filaments. 
Work along these lines is in progress.

\begin{acknowledgments}
The authors wish to thank M. Baus, C. Pierleoni and G. Ciccotti for useful 
discussions about the present work. 
J.-P. R. hopes that the honoured and friend G. Ciccotti will appreciate the 
present work, despite the marginal use of Lagrange multipliers and the 
audacious choice of the infinitely stiff bond model for the discrete 
wormlike chain, avoiding the straight imposition of holonomic rigid bond 
constraints to enforce the filament fixed contour length property and 
therefore disregarding the beloved Jacobians!
The authors warmly thank G. Destr\'{e}e for invaluable technical help. 
S. Ramachandran acknowledges financial help from the BRIC Department 
(Bureau des Relations Internationales et de Coop\'eration) of 
the Universit\'e Libre de Bruxelles. 
\end{acknowledgments}


\nocite{*}
 \newcommand{\noopsort}[1]{} \newcommand{\printfirst}[2]{#1}
  \newcommand{\singleletter}[1]{#1} \newcommand{\switchargs}[2]{#2#1}
%

\end{document}